\documentclass[letterpaper, 10 pt, conference]{ieeeconf}  

\usepackage{graphics} 
\usepackage{subfigure}
\usepackage{epsfig} 
\usepackage{color} 
\usepackage{url}
\usepackage{cite}
\usepackage{algorithm, algorithmic}
\usepackage{multicol}
\usepackage{amssymb,dsfont,amsmath,amsfonts,mathrsfs,soul}
\usepackage{xspace,colortbl,graphicx}
\usepackage{enumerate} 
\usepackage{mathtools} 
\usepackage{color, soul} 
\usepackage{multirow}
\usepackage{mathtools} 

\usepackage{cases}

\sethlcolor{yellow}

\IEEEoverridecommandlockouts                              

\usepackage{cite}
\title{\LARGE \bf
Assessing the safety benefits of CACC+ based coordination of connected and autonomous vehicle platoons in emergency braking scenarios}

\author{Guoqi Ma, Prabhakar R. Pagilla$^\ast$, Swaroop Darbha
\thanks{The authors are with the Department of Mechanical Engineering, Texas A\&M University, College Station, TX 77843, USA. E-mails: 
 {\tt \{gqma, ppagilla, dswaroop\}@tamu.edu}.
 }
 \thanks{$^\ast$Corresponding author.
 }
}

\begin{document}

\maketitle
\thispagestyle{empty}
\pagestyle{empty}

\begin{abstract}
Ensuring safety is the most important factor in connected and autonomous vehicles, especially in emergency braking situations. As such, assessing the safety benefits of one information topology over other is a necessary step towards evaluating and ensuring safety. In this paper, we compare the safety benefits of a cooperative adaptive cruise control which utilizes information from one predecessor vehicle (CACC) with the one that utilizes information from multiple predecessors (CACC+) for the maintenance of spacing under an emergency braking scenario. A constant time headway policy is employed for maintenance of spacing (that includes a desired standstill spacing distance and a velocity dependent spacing distance) between the vehicles in the platoon. The considered emergency braking scenario consists of braking of the leader vehicle of the platoon at its maximum deceleration and that of the following vehicles to maintain the spacing as per CACC or CACC+. By focusing on the standstill spacing distance and utilizing Monte Carlo simulations, we assess the safety benefits of CACC+ over CACC by utilizing the following safety metrics: (1) probability of collision, (2) expected number of collisions, and (3) severity of collision (defined as the relative velocity of the two vehicles at impact). We present and provide discussion of these results. 

\end{abstract}

\section{INTRODUCTION}
 Connected and autonomous vehicles (CAVs) have been receiving extensive attention in recent decades due to the substantial benefits they could offer in terms of increasing traffic capacity, improving fuel economy and reducing  transportation costs~\cite{darbha2018benefits,mouftah2020connected,papadoulis2019evaluating}. Utilization of advanced communication technologies, such as Dedicated Short Range Communications (DSRC)~\cite{5888501}, Vehicle-to-Vehicle (V2V) communication~\cite{vinel2015vehicle}, or more inclusive Vehicle-to-Everything (V2X) communication~\cite{9345798}, can be a strong enabler for CAVs to achieve such benefits. Most existing studies have focused on string stability of the platoon in the presence of external disturbances, i.e., the effect of information topology on the inter-vehicular spacing error propagation along the vehicle stream, see  e.g.~\cite{iet2020platooning,vegamoor2021string,6683051,6515636,7879221,MALIKOPOULOS2021109469,FENG201981,BIAN201987,GE2018445,10.1115/1.4036565,KIM2022110488} and references therein. While string stability can ensure collision avoidance, however, such performance is usually achieved without considering of the capabilities of each vehicle in the platoon, i.e.,  actuator saturation or acceleration/deceleration limitations of the vehicles. 



A significant safety situation could occur due to an emergency braking scenario, where the lead vehicle may have to brake with its maximum deceleration to avoid a crash. Ideally, if all the following vehicles can brake with the same deceleration as the lead vehicle and react instantaneously, then there would be no collisions. However, in practice, the maximum deceleration capacity of each vehicle is not identical for many reasons (wear, type of vehicle, road conditions, etc.,)~\cite{godbole1997tools}.
Although the following vehicles could brake with their own maximum deceleration, this could cause exacerbated collision occurrence when the maximum deceleration is non-identical for the vehicles in the platoon. Due to the complexity of the factors that influence safety, such as uncertain/stochastic actuator saturation, initial velocity, standstill spacing, control gain values, it is difficult to analytically derive the safety metrics for quantitative evaluations related to these factors. Therefore, a realistic, yet beneficial, way to assess the safety metrics is by using numerical methods. Further, it is also infeasible to utilize a brute force approach to numerically consider and evaluate the safety metrics for all the possible combinations in the search space. As a result, a practical and tractable way is to estimate the safety metrics through empirical simulation algorithms, such as the Monte Carlo simulation approach, by considering some key aspects such as standstill spacing, time headway and the number of predecessor vehicles. Such numerical strategies have been employed to draw meaningful conclusions on safety in the previous work~\cite{vegamoor2019review,darbha2012methodology,darbha2013methodology,choi2001assessingACC,choi2001assessingITSC}. 


    
However, existing work focused on the CACC case where only information from the immediate predecessor was utilized. Motivated by the work in~\cite{darbha2018benefits} which highlighted the benefits of V2V communication on improving platoon tightness and mobility, in this paper we highlight the safety benefits of CACC+ based CAVs where multiple-predecessor vehicles' information is communicated on improving the aforementioned safety metrics in emergency braking scenarios. By assuming the vehicles have an independent and identically distributed (i.i.d.) maximum deceleration, we will consider the case that the lead vehicle brakes with its maximum deceleration while the following vehicles brake according to the designed coordination control law subject to a stochastically distributed maximum deceleration. We will evaluate the safety metrics under all the possible maximum deceleration values of the lead vehicle. In addition, we will choose one of the influencing factors on safety metrics, the standstill spacing, to evaluate the effectiveness of CACC+ under a smaller standstill spacing in ameliorating the safety metrics by comparing with CACC under a larger standstill spacing. The main contributions of the paper lie in problem setup and formulation, including the modeling of the emergency scenario and simulation framework, which can be utilized by the ITS community to evaluate safety metrics on various CAVs with various information topologies.        
 



The remainder of the paper is organized as follows. Section~\ref{section:problem-formulation-and-preliminaries} contains preliminaries including the vehicle dynamics and control law associated with CACC and CACC+ with the constant time headway policy. The simulation algorithm is  developed in Section~\ref{section:main-results}. An illustrative numerical example and simulation results are provided in Section~\ref{section:numerical-simulations}. Finally, some concluding remarks are given in Section~\ref{section:conclusion}.


\section{PRELIMINARIES} \label{section:problem-formulation-and-preliminaries}
\subsection{Vehicle String Model}
\begin{figure}[!htb]
\centering{\includegraphics[scale=0.55]{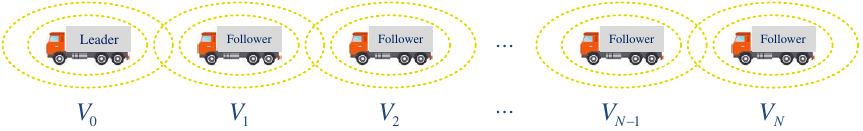}}
\caption{An illustration of the connected and autonomous vehicle platoon with V2V communication.
\label{fig_illustration_of_vehicle_platoon}}
\end{figure}

Consider a string of homogeneous vehicles equipped with V2V communication as illustrated in Fig.~\ref{fig_illustration_of_vehicle_platoon}.
The dynamics model of the $i$-th vehicle in the string is given by
\begin{align} \label{eq:vehicle-dynamics}
\left\{\begin{array}{*{20}{l}}
\dot{x}_{i}(t) = v_{i}(t), \\
\dot{v}_i(t) = a_i(t), \\
\tau \dot{a}_{i}(t) + a_{i}(t) = u_{i}(t),
\end{array}
\right.
\end{align}
where $x_{i}(t)$, $v_i(t)$, $a_{i}(t)$, $u_{i}(t)$ represent the position, velocity, acceleration, and control input of the $i$-th vehicle at time instant $t$, and $i \in \mathcal{N} = \{ 1, 2, \cdots, N \}$, where $N$ is the total number of the following vehicles in the platoon, $\tau$ denotes the parasitic actuation lag. It is assumed that $\tau$ is {\it uncertain} with $\tau \in \left( 0, \tau_{0} \right]$, where $\tau_{0}$ is a positive real constant. 

We first define the generalized or velocity-dependent inter-vehicular spacing error for the $i$-th vehicle as follows: 
\begin{align} \label{eq:delta-i-definition}
 \delta_i(t) = e_i(t) + h_w v_i(t),   
\end{align}
where $e_i(t) = x_{i}(t) - x_{i - 1}(t) + d$ is the spacing error for the $i$-th vehicle, where $d$ is the minimum or standstill spacing between adjacent vehicles, 
$h_w$ is the time headway. 





\subsection{Coordination Control Law Design}
Consider the following decentralized coordination control law for the $i$-th vehicle:
\begin{align} \label{eq:continuous-time-u-i}
u_i(t) &= \sum\limits_{q = 1}^r \Big( k_a a_{i - q}(t) -  k_v ( v_i(t) - v_{i - q}(t) ) \nonumber \\
& ~ ~ - k_p ( x_i(t) - x_{i - q}(t) + d q + q h_w v_i(t) ) \Big),
\end{align}
where $k_a$, $k_v$, $k_p$ are positive controller gains to be designed, $r$ denotes the number of ``look-ahead" predecessors for vehicle $i$. Then the inter-vehicular spacing error propagation equation is given by
\begin{align}
 \delta_i(s) = H(s) \sum\limits_{q = 1}^{r} \delta_{i - q}(s), 
\end{align}
 where $H(s) = N_H(s)/D_H(s)$ with $N_H(s) = k_a s^2 + k_v s + k_p$, $D_H(s) = \tau s^3 + s^2 + \gamma s + r k_p$ with 
 $\gamma = r k_v + r k_p \frac{r + 1}{2} h_w$. As discussed in~\cite{darbha2018benefits}, a sufficient condition to ensure robust string stability (robustness with respect to the parasitic actuation lag, $\tau\in (0,\tau_0]$) of the vehicle platoon is given by 
\begin{align}
 \Vert r H(s) \Vert_{\infty} \le 1. 
\end{align}
Let $\tilde{H}(s) = r H(s)$,
and define $\tilde{k}_a = r k_a$, $\tilde{k}_v = r k_v$, $\tilde{k}_p = r k_p$, $\tilde{h}_w = \frac{r + 1}{2} h_w$, then 
 from $\Vert \tilde{H}(j \omega) \Vert_{\infty} \le 1$, we can obtain $\tilde{k}_a \in (0,1)$ and the admissible region of $\tilde{k}_v$ and $\tilde{k}_p$ as 
\begin{align} \label{eq:tilde-kv-tilde-kp-admissible-region-1}
\begin{cases}{}
 \frac{ \tilde{k}_v }{ a_1 } + \frac{ \tilde{k}_p }{ b_1 } \le 1, \\
 \frac{ \tilde{k}_v }{ a_2 } + \frac{ \tilde{k}_p }{ b_2 } \ge 1,
\end{cases}
\end{align}
 where 
\begin{align*}
 a_1 = \frac{ 1 - \tilde{k}_a^2 }{ 2 \tau_0 }, b_1 = \frac{ 1 - \tilde{k}_a^2 }{ 2 \tau_0 \tilde{h}_w }, a_2 = \frac{ 1 - \tilde{k}_a }{ \tilde{h}_w }, b_2 = \frac{ 2 ( 1 - \tilde{k}_a ) }{ \tilde{h}_w^2 }.
\end{align*}
 In order to ensure~\eqref{eq:tilde-kv-tilde-kp-admissible-region-1} has positive solutions for $\tilde{k}_v$ and $\tilde{k}_p$, we need $a_1 > a_2$, i.e., 
\begin{align}
 \frac{ 1 - \tilde{k}_a^2 }{ 2 \tau_0 } > \frac{ 1 - \tilde{k}_a }{ \tilde{h}_w },
\end{align}   
 from which we can obtain $\tilde{h}_w > \frac{ 2 \tau_0 }{ 1 + \tilde{k}_a }$, that is, $h_w$ satisfies: 
\begin{align}
 \tilde{h}_w > \frac{ 4 \tau_0 }{(1+r)(1 + r {k}_a)}. 
\end{align}

\begin{figure}[!htb]
\centering{\includegraphics[scale=0.65]{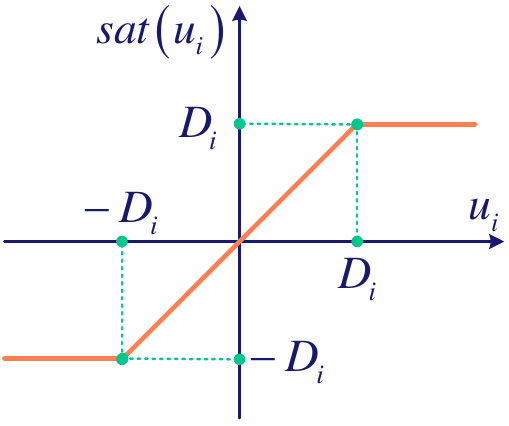}}
\caption{An illustration of the saturation function $sat(u_i)$.
\label{fig_saturation_function}}
\end{figure}

Note that the control law~\eqref{eq:continuous-time-u-i} does not take into account actuator saturation, or vehicle acceleration/deceleration capability. When this constraint is incorporated, the actual input to the $i$-th vehicle, denoted as $\bar{u}_i$, is given by
\begin{align} \label{eq:saturation_function}
\bar{u}_i = sat(u_i) =  
\begin{cases}{}
 -D_i, \ \mbox{when}~u_i < -D_i, \\
  u_i, \ \ \ \ \  \mbox{when}~-D_i \le u_i \le D_i, \\
 D_i, \ \ \ \ \mbox{when}~u_i > D_i. 
 \end{cases}
\end{align}
An illustration of the saturation function $sat(u_i)$ is shown in Fig.~\ref{fig_saturation_function}.

In practice, it is reasonable to assume the vehicle maximum acceleration/deceleration, $D_i$, varies in a bounded interval such that $D_i \in [D_{\rm lower}, D_{\rm upper}]$. Besides, to facilitate numerical evaluations in this range of $D_i$, we model the interval $[D_{\rm lower}, D_{\rm upper}]$ as arithmetic progression with a certain probability distribution, i.e., 
\begin{align}
\begin{cases}{}
 D_i \in \{ \hat{D}_1, \hat{D}_2, \cdots, \hat{D}_m \}, \\
 \mathbb{P}\{ D_i = \hat{D}_j \} = p_j, j = 1, \cdots, m, \\
 \sum\limits_{j = 1}^m p_j = 1,
 \end{cases}
\end{align}
where $\hat{D}_1 = D_{\rm lower}$, $\hat{D}_m = D_{\rm upper}$, $\mathbb{P}\{ D_i = \hat{D}_j \}$ denotes the probability of $D_i$ taking the value of $\hat{D}_j$, and $m$ is the discretization number of the interval $[D_{\rm lower}, D_{\rm upper}]$.

\section{Setup and Approach for Assessing Safety} \label{section:main-results}




 \label{Section_Main_Results}


\subsection{Discretization of the Continuous-Time Dynamics Model}

To discrete the continuous-time vehicle dynamics for implementation in computer simulation, we consider $h$ to be the discretization time step size, $k$ be the discretization index, and  $x_{i}[k], v_{i}[k], a_{i}[k]$ denote the discretized position, velocity, and acceleration of the $i$-th vehicle at time step $k$. Let the initial conditions for the $i$-th vehicle be: $x_i[0] = \bar{x}_i = - i d - i h_w v_i[0]$, $v_i[0] = \bar{v}_i$, $a_i[0] = 0$, where $\bar{v}_i$ is the steady-state velocity. By utilizing the fourth-order Runge-Kutta approach, the continuous-time dynamics for the $i$-th vehicle as given by~\eqref{eq:vehicle-dynamics} can be discretized as follows:
 \begin{align} \label{eq:runge-kutta-equation}
 \begin{cases}{}
 x_i[k + 1] = x_i[k] + v_i[k] h, \\
 v_i[k + 1] = v_i[k] + a_i[k] h, \\
 a_i[k + 1] = a_i[k] + \frac{h}{6} ( k_1 + 2 k_2 + 2 k_3 + k_4 ),
 \end{cases}
\end{align}
 where  
 \begin{align}
 \begin{cases}{}
 k_1 = -\frac{1}{\tau} a_i[k] + \frac{1}{\tau} u_i[k],  \\
 k_2 = -\frac{1}{\tau} ( a_i[k] + \frac{h}{2} k_1 ) + \frac{1}{\tau} u_i[k], \\
 k_3 = - \frac{1}{\tau} ( a_i[k] + \frac{h}{2} k_2 ) + \frac{1}{\tau} u_i[k], \\
 k_4 = - \frac{1}{\tau} ( a_i[k] + h k_3 ) + \frac{1}{\tau} u_i[k], 
 \end{cases}
 \end{align}
and $u_i[k]$ is the discretized version of~\eqref{eq:continuous-time-u-i}, which is given by
\begin{align}
u_i[k] &= \sum\limits_{q = 1}^r \Big( k_a a_{i - q}[k] -  k_v ( v_i[k] - v_{i - q}[k] ) \nonumber \\
& ~ ~ - k_p ( x_i[k] - x_{i - q}[k] + d q + q h_w v_i[k] ) \Big).
\end{align}

For the first $(r - 1)$ following vehicles that have less than $r$ predecessor vehicles for communication, e.g., vehicle 1 and vehicle 2 when $r = 3$, the control law is given by
\begin{align}
 u_i[k] &= \sum\limits_{q = 1}^i \Big( k_a a_{i - q}[k] - k_v ( v_i[k] - v_{i - q}[k] ) \nonumber \\
& ~ ~ - k_p ( x_i[k] - x_{i - q}[k] + d q + q h_w v_i[k] \Big). 
\end{align}
Taking actuator saturation into account, the actual control input to the $i$-th vehicle at time step $k$ is given by
\begin{align}
 \bar{u}_i[k] = sat(u_i[k]),
\end{align}
where the saturation function $sat(.)$ is defined in~\eqref{eq:saturation_function}. 

\subsection{Monte Carlo Simulation Approach}

 Monte Carlo simulation is an effective way to estimate the expected value of a random variable. According to~\cite{hoeffding1994probability}, the accuracy of the estimation is quantified by
\begin{align}
 Prob\{ \vert X_n - X_t \vert > \varepsilon \} \le 2 e^{-2 n \varepsilon^2}, 
\end{align}
where $X_n$ and $X_t$ represent the mean estimated value after $n$ iterations and the true expected value for a random variable $X$, respectively. Then, we have  
\begin{align}
 \lim\limits_{n \rightarrow \infty} Prob\{ \vert X_n - X_t \vert > 0 \} = 0,
\end{align} 
i.e. $\lim\limits_{n \rightarrow \infty} X_n = X_t$. Thus, for a sufficiently large $n$, the estimated expected value can tend to the true expected value of the random variable. As such, in the following, we will adopt the Monte Carlo simulation approach to estimate the expected values of the  safety metrics.   

\subsection{Collision Model and Safety Metrics}
\subsubsection{Collision Model}
For each iteration, at time step $k$, if
\begin{align}
 x_i[k] \ge x_{i - 1}[k],
\end{align}
then a collision occurs between vehicle $i$ and vehicle $i - 1$ at time step $k$. If a collision occurs between vehicle $i$ and vehicle $i + 1$, without considering the coefficient of restitution~\cite{darbha2012methodology}, we assume that vehicle $i$ and vehicle $i - 1$ will stop immediately and will not move any more, i.e., 
\begin{align}
\begin{cases}{}
 a_i[l] = a_{i - 1}[l] = 0, \\
 v_i[l] = v_{i - 1}[l] = 0, \\
 x_i[l] = x_i[k], x_{i - 1}[l] = x_{i - 1}[k],
\end{cases}
\end{align} 
$l = k + 1, k + 2, \cdots, K$, where $K$ is the total number of simulation steps.   

\subsubsection{Safety Metrics} We will consider the following safety metrics: (1) the probability of collision; (2) the expected number of collisions; and (3) the severity of collisions (defined as the relative velocity of collisions at impact). 

First, the probability of collision $\mathbb{P}$ is defined as 
 \begin{align}
 \mathbb{P} = \frac{\mathcal{C}}{n} \times 100\%,
 \end{align}
where $\mathcal{C} = \sum\limits_{j = 1}^n C(j)$, $n$ is the total iterations (or samples), and the binary number $C(j)$ indicates whether there is a collision in the $j$-th iteration and is given by 
\begin{align}
 \begin{cases}{}
 C(j) = 0, \mbox{if there is no collision in the $j$-th iteration}; \\
C(j) = 1, \mbox{if there is a collision in the $j$-th iteration}.
 \end{cases}
\end{align} 

Second, the expected number of collisions $\mathbb{N}$ is defined as
\begin{align}
 \mathbb{N} = \frac{\mathcal{N}}{n},
\end{align}
where $\mathcal{N} = \sum\limits_{j = 1}^n N(j)$, where $N(j)$ denotes the number of collisions in the $j$-th iteration. 

Third, the severity of collision is defined as 
\begin{align}
 \mathbb{S} = \frac{\mathcal{V}}{n},
\end{align}
where $\mathcal{V} = \sum\limits_{j = 1}^n V(j)$ and $V(j)$ denoting the sum of the relative velocity of collisions in the $j$-th iteration.

Based on the above definitions of the collision model and safety metrics, the simulation procedure employed is provided in~{\bf Algorithm~\ref{Algorithm_1}}.

\begin{algorithm}[h!]
\caption{Monte Carlo Safety Metric Evaluation Algorithm for Connected and Autonomous Vehicle Platoons with i.i.d. Maximum Deceleration in Emergency Braking Scenarios}
\begin{algorithmic}[1] \label{Algorithm_1}
\STATE Initialization: Given platoon length $N$, initial values for vehicles $a_l[0] = 0$, $v_l[0] = \bar{v}_l$, $x_l[0] = \bar{x}_l$, $l = 1, \cdots, N$, the number of predecessors providing information $r$, the control gains  $k_a$, $k_v$, $k_p$, $h_w$ satisfying robust string stability condition, set maximum number of iteration steps $n$, discretization time step $h$, simulation time $\mathscr{T}$, and the number of simulation steps $K = \mathscr{T}/h$; and set collision number $CN = 0$. Given the i.i.d. maximum deceleration set $\mathscr{D}$, generate a probability matrix $\mathcal{P}_{n \times N} = {\rm unifrnd}(0, 1, [n, N])$, and the maximum deceleration matrix $\mathcal{D}_{n \times N}$ according to $\mathscr{D}$ and $\mathcal{P}_{n \times N}$.  
\FOR{each $i \in [1, n]$}
\STATE Let $CS_l = 0$, $RV_l = 0$, $D_l = \mathcal{D}(i, l)$, $l = 1, \cdots, N$; \% $CS_l$ denotes the collision status of the $l$-th vehicle, $RV_l$ denotes the relative velocity at impact for the $l$-th vehicle with its predecessor vehicle, $D_l$ denotes the maximum deceleration of the $l$-th vehicle; \label{Algorithm_fitness}
\FOR{$j \in [1, K]$ }
\STATE \textbf{if} $CS_l \cup CS_{l + 1} == 1$ \textbf{then} \\
 $a_l(j + 1) = 0$, $v_l(j + 1) = 0$, $x_l(j + 1) = x_l(j)$; \\ \textbf{else} perform the fourth-order Runge-Kutta discretization of the vehicle dynamics according to~\eqref{eq:runge-kutta-equation}; \\
 \textbf{end if}
\STATE \textbf{if} {$x_l[j + 1] \ge x_{l - 1}[j + 1]$ \& $CS_l == 0$} \textbf{then} {$CS_l = 1$, $RV_l = v_l[j + 1] - v_{l - 1}[j + 1]$};
\STATE \textbf{end if}
\ENDFOR
\STATE \textbf{if} {$\bigcup\limits_{l = 1, \cdots, N}CS_l == 1$} \textbf{then} {$CN++$};
\STATE \textbf{end if}
\STATE Compute $CS(i) = \sum\limits_{l = 1}^N CS_l$; $RV(i) = \sum\limits_{l = 1}^N RV_l$;
\STATE \textbf{if} {$CS(i) \ge 1$} \textbf{then} compute {$severity(i) = RV(i)/CS(i)$}; 
\STATE \textbf{end if}
\ENDFOR
\STATE Output the probability of collision as $\mathbb{P} = CN/n$, the expected number of collisions as $\mathbb{N} = \sum\limits_{i = 1}^n CS(i)/n$, and the relative velocity at impact as $\mathbb{S} = \sum\limits_{i = 1}^n severity(i)/n$. \label{Algorithm_Step_Last}
\end{algorithmic}
\end{algorithm}


\section{SIMULATION RESULTS AND DISCUSSIONS} \label{section:numerical-simulations}
\begin{figure}[!htb]
\centering{\includegraphics[scale=0.55]{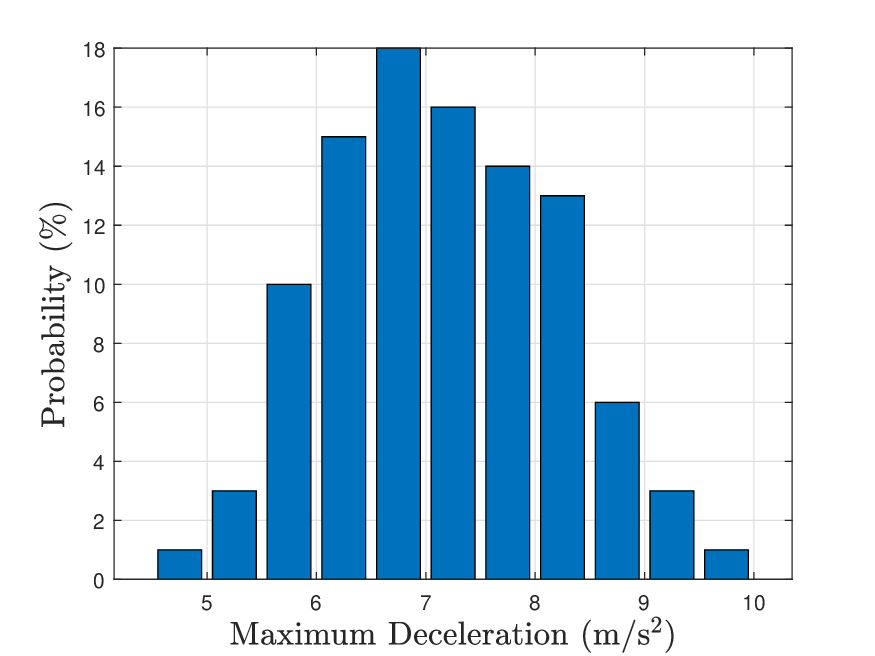}}
\caption{The probability distribution of the maximum deceleration~\cite{godbole1997tools}.
\label{fig_maximum_deceleration}}
\end{figure}

In this section, we present a numerical example to perform the simulation algorithm achieved in Section~\ref{section:main-results}. We consider the following numerical values for the system parameters: $N = 10$, $\tau_0 = 0.5$ seconds, $d= 2, 4, 6$ m, respectively. In the numerical simulation, $\tau$ was chosen as $\tau = \tau_0$. The initial steady-state velocity for the $i$-th vehicle is assumed to be $\bar{v}_i = 25~{\rm m/s}$. The number of iterations is chosen to be $n = 2000$, and the simulation time as $50$ seconds with the discretization step $h = 0.01$ seconds. 
We consider the maximum deceleration of the $i$-th vehicle $D_i \in [4.75, 9.75]$ with the probability distribution as given in Fig.~\ref{fig_maximum_deceleration}.  



\begin{figure}[!htb]
\centering{\includegraphics[scale=0.55]{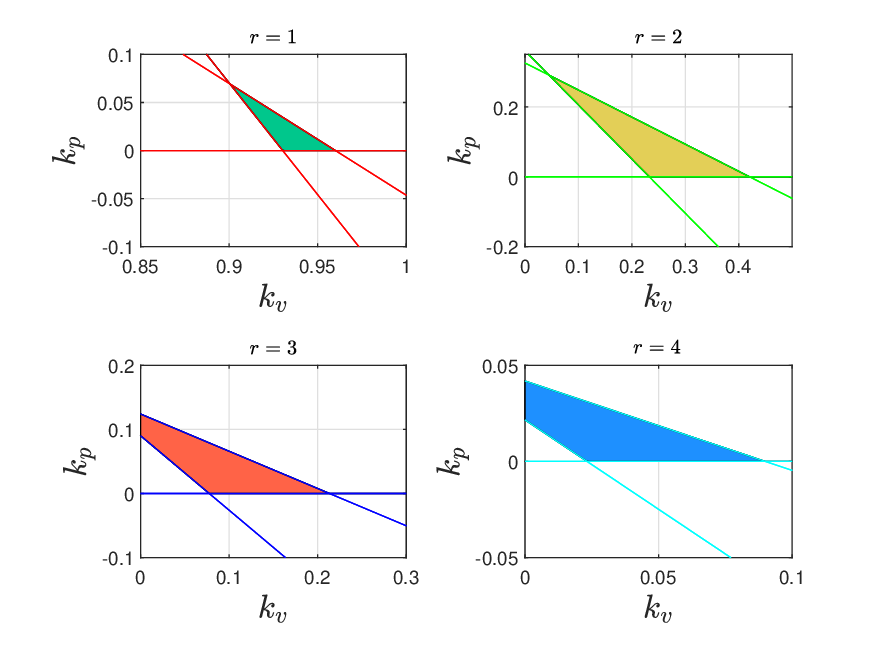}}
\caption{The admissible regions of $k_v$ and $k_p$.
\label{fig:kvkp}}
\end{figure}

Choose $k_a = 0.2$, $h_w = 0.86$, then the admissible regions of $k_v$ and $k_p$ for $r = 1, 2, 3, 4$ are shown in Fig.~\ref{fig:kvkp}.
In an emergency braking scenario, collision avoidance is a key objective. Thus, we choose the control gains and time headway for the CACC+ case ($r>1$) to be the same as that of the CACC case ($r=1$) with $k_v = 0.92$, $k_p = 0.03$, guaranteeing a fair comparison between CACC and CACC+. Before proceeding with this, safety metrics for the case without coordination, i.e., all vehicles brake with the maximum deceleration, are provided in TABLE~\ref{table:safety-metrics-without-coordination}. Note that collision avoidance without coordination may be ensured only in limited cases such as $D_0 < D_1 < D_2 < ... < D_{10}$; since we have 11 discretized values of $D_i$, the probability of collision avoidance in this case is near zero; to see this, let $p_1 = \mathbb{P} \{ D_0 = \hat{D}_1 \}$, $p_2 = \mathbb{P} \{ D_1 = \hat{D}_2 \}$, $\cdots$; since the set $[\hat{D}_1, \cdots, \hat{D}_{11}]$ has 11 values, the probability of collision avoidance without coordination is given by $\mathbb{P}\{ D_0 < D_1 < ... < D_{10} \} = \prod_{j = 1}^{11} p_j$, which is near zero. Therefore, the probability of collision without coordination is almost 1. 

\begin{table*}[ht!]
\vspace*{0.1in}
\centering
\caption{The safety metrics for the case without coordination ($d = 6~{\rm m}$)}
\label{table:safety-metrics-without-coordination}
\begin{tabular}{|c|c|c|c|c|c|c|c|c|c|c|c|}
\hline
$D_0$ & 4.75 & 5.25 & 5.75 & 6.25 & 6.75 & 7.25 & 7.75 & 8.25 & 8.75 & 9.25 & 9.75 \\
 \hline
$\mathbb{P}$ & 1 & 1 & 1 & 1 & 1 & 1 & 1 & 1 & 1 & 1 & 1 \\
\hline
$\mathbb{N}$ & 6.1680 & 6.2405 & 6.2080 & 6.2150 & 6.2085 & 6.1970 & 6.2345 & 6.2440 & 6.2125 & 6.1940 & 6.2350 \\
\hline
$\mathbb{S}$ & 22.5985 & 22.5913 & 22.4154 & 22.7564 & 22.6850 & 22.7315 & 22.8060 & 22.8738 & 22.7030 & 22.5595 & 22.9056 \\
\hline 
\end{tabular}
\end{table*}

To demonstrate the effectiveness of coordination in reducing the three safety metrics, we consider the following cases $r = 1$ (CACC), $r = 2$ (CACC+), and $r = 3$ (CACC+) with three different standstill spacing $d=2$ m, $d=4$ m, and $d=6$ m.

\begin{figure}[!htb]
\centering{\includegraphics[scale=0.55]{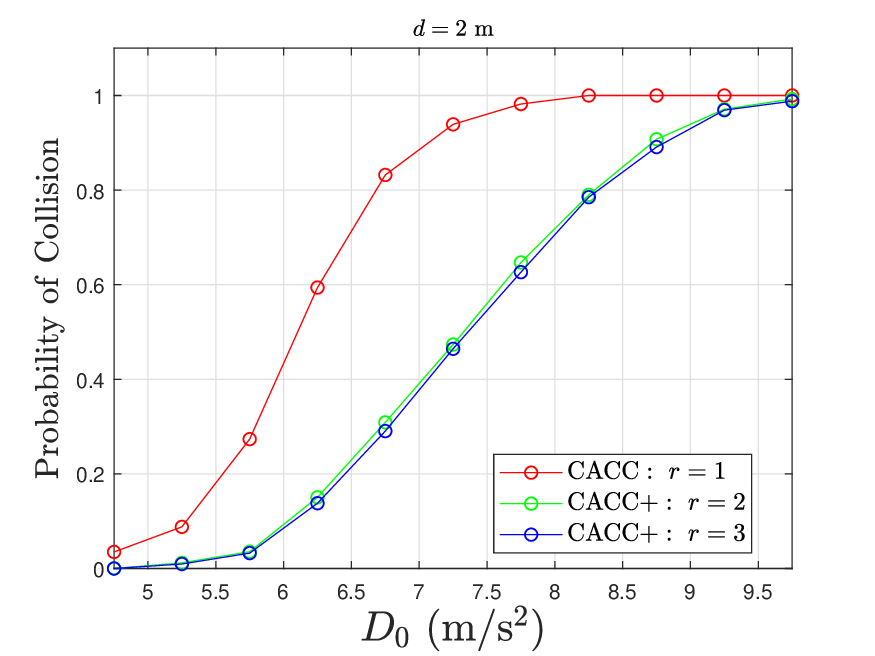}}
\caption{The probability of collision when $d = 2$ m.
\label{fig:P_d_2}}
\end{figure}

\begin{figure}[!htb]
\centering{\includegraphics[scale=0.55]{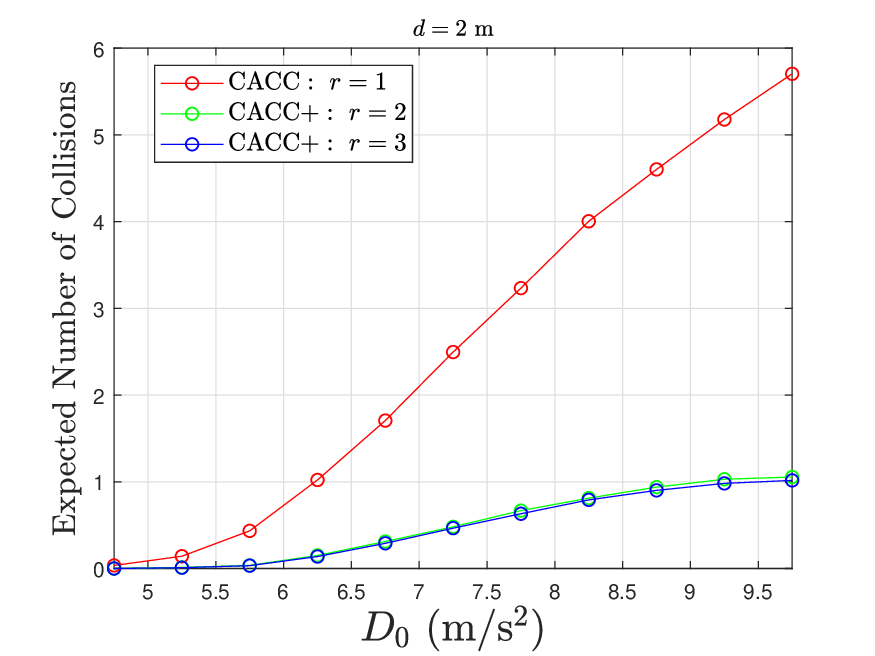}}
\caption{The expected number of collisions when $d = 2$ m.
\label{fig:N_d_2}}
\end{figure}

\begin{figure}[!htb]
\centering{\includegraphics[scale=0.55]{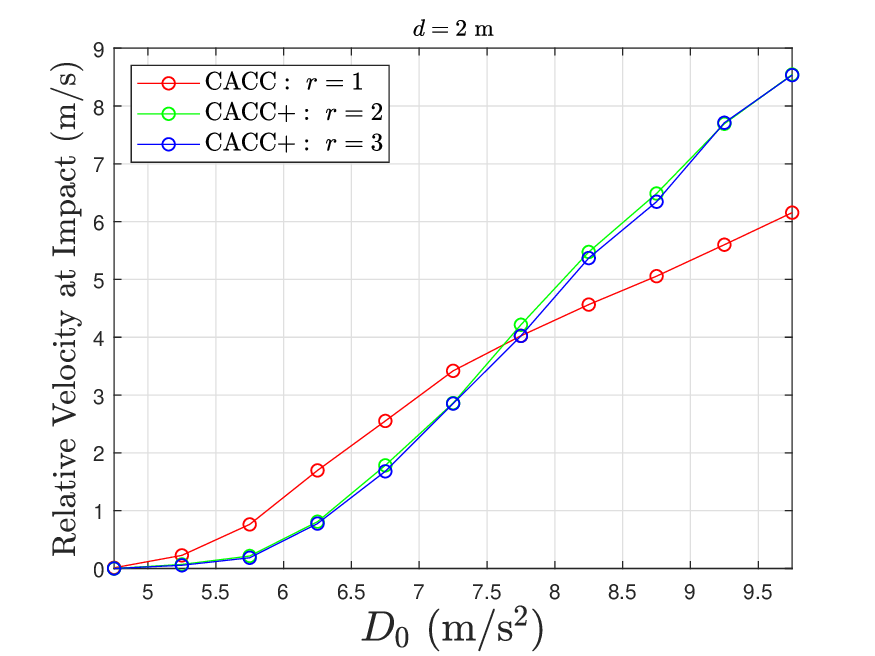}}
\caption{The relative velocity at impact when $d = 2$ m.
\label{fig:S_d_2}}
\end{figure}

First, when the standstill spacing $d = 2$ m, the probability of collision, the expected number of collisions, and the relative velocity at impact are provided in Fig.~\ref{fig:P_d_2},~Fig.~\ref{fig:N_d_2}, and~Fig.~\ref{fig:S_d_2}, respectively.

\vspace*{-0.1in}

\begin{figure}[!htb]
\centering{\includegraphics[scale=0.55]{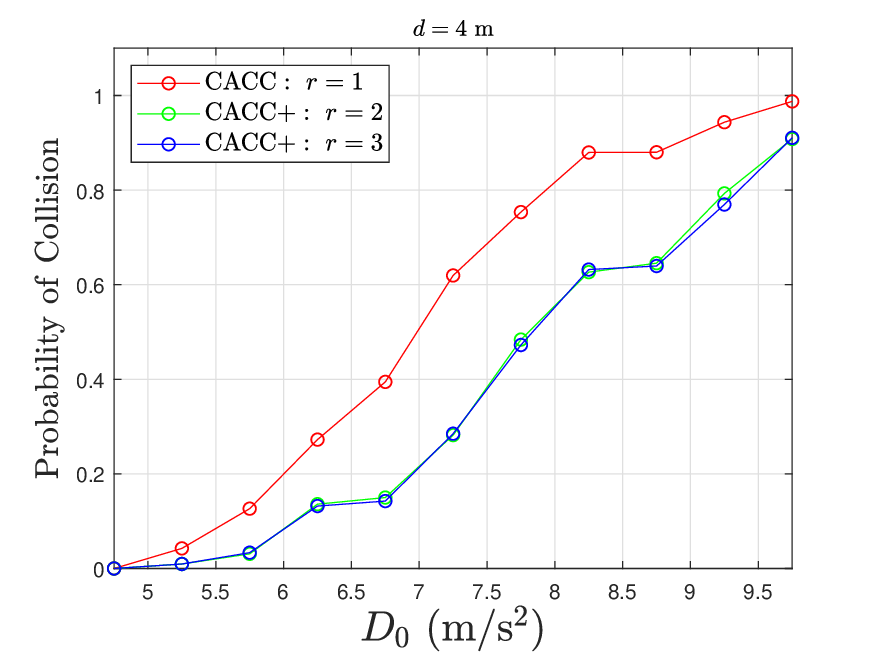}}
\caption{The probability of collision when $d = 4$ m.
\label{fig:P_d_4}}
\end{figure}

\begin{figure}[!htb]
\centering{\includegraphics[scale=0.55]{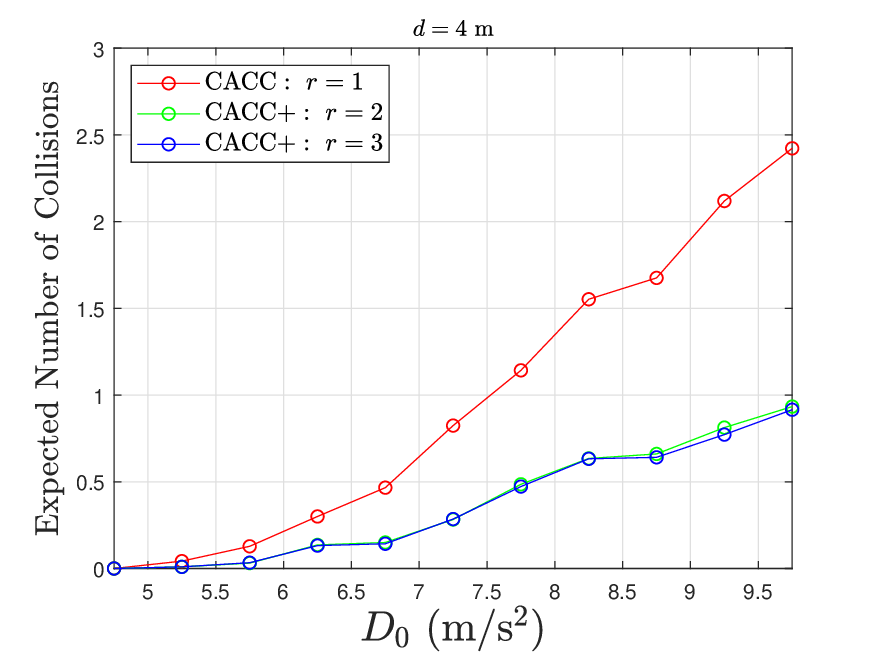}}
\caption{The expected number of collisions when $d = 4$ m.
\label{fig:N_d_4}}
\end{figure}

\begin{figure}[!htb]
\centering{\includegraphics[scale=0.55]{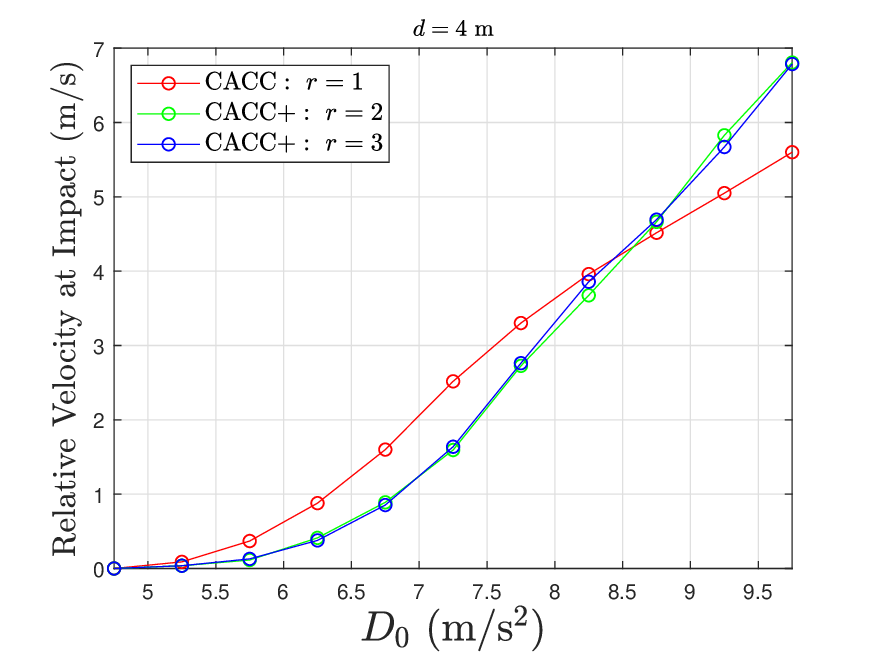}}
\caption{The relative velocity at impact when $d = 4$ m.
\label{fig:S_d_4}}
\end{figure}

Second, when the standstill spacing $d = 4$ m, the probability of collision, the expected number of collisions, and the relative velocity at impact, respectively, are provided in Fig.~\ref{fig:P_d_4},~Fig.~\ref{fig:N_d_4}, and~Fig.~\ref{fig:S_d_4}. 

\begin{figure}[!htb]
\centering{\includegraphics[scale=0.55]{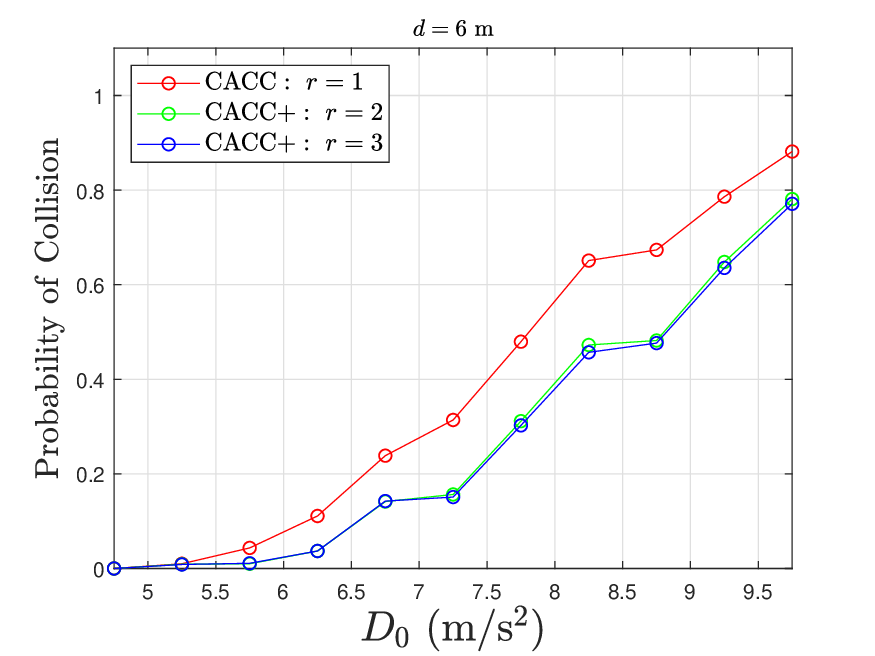}}
\caption{The probability of collision when $d = 6$ m.
\label{fig:P_d_6}}
\end{figure}

\begin{figure}[!htb]
\centering{\includegraphics[scale=0.55]{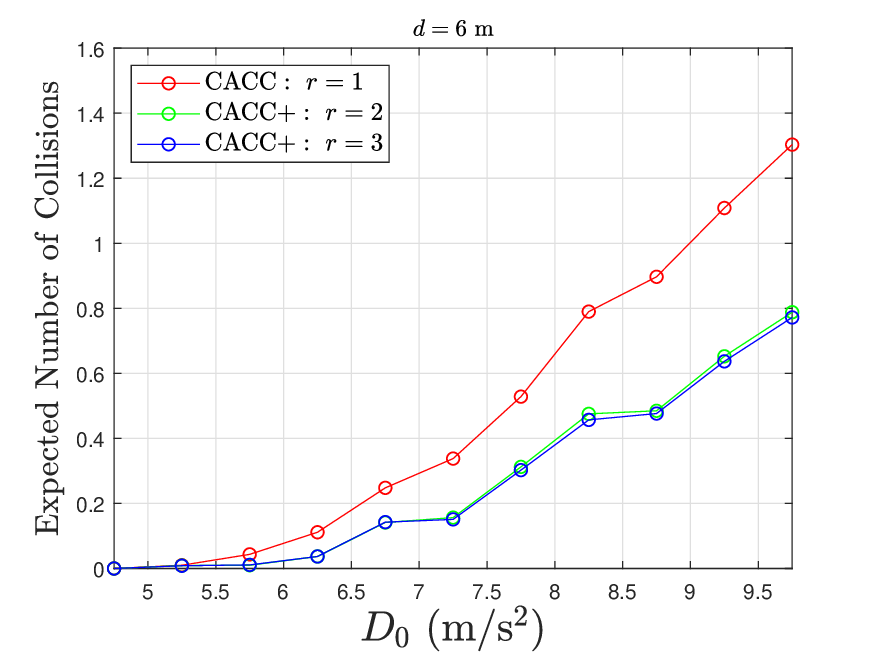}}
\caption{The expected number of collisions when $d = 6$ m.
\label{fig:N_d_6}}
\end{figure}

\begin{figure}[!htb]
\centering{\includegraphics[scale=0.55]{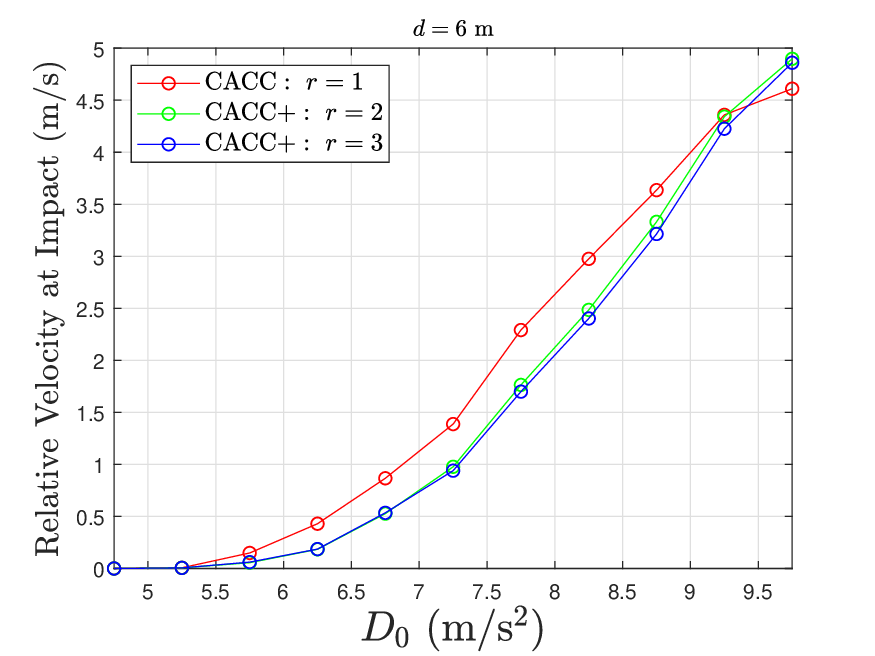}}
\caption{The relative velocity at impact when $d = 6$ m.
\label{fig:S_d_6}}
\end{figure}

In addition, when the standstill spacing $d = 6$ m, the probability of collision, the expected number of collisions, and the relative velocity at impact, respectively, are provided in Fig.~\ref{fig:P_d_6},~Fig.~\ref{fig:N_d_6}, and~Fig.~\ref{fig:S_d_6}. 

\vspace*{-0.1in}

From the above simulation results, one can observe that the probability of collision and the expected number of collisions can be improved under CACC+, whereas the relative velocity at impact can deteriorate when $D_0$ is greater than a certain value, which may be due to several reasons, such as the heterogeneous control structure, parameter values, etc. 

\begin{figure}[!htb]
\centering{\includegraphics[scale=0.55]{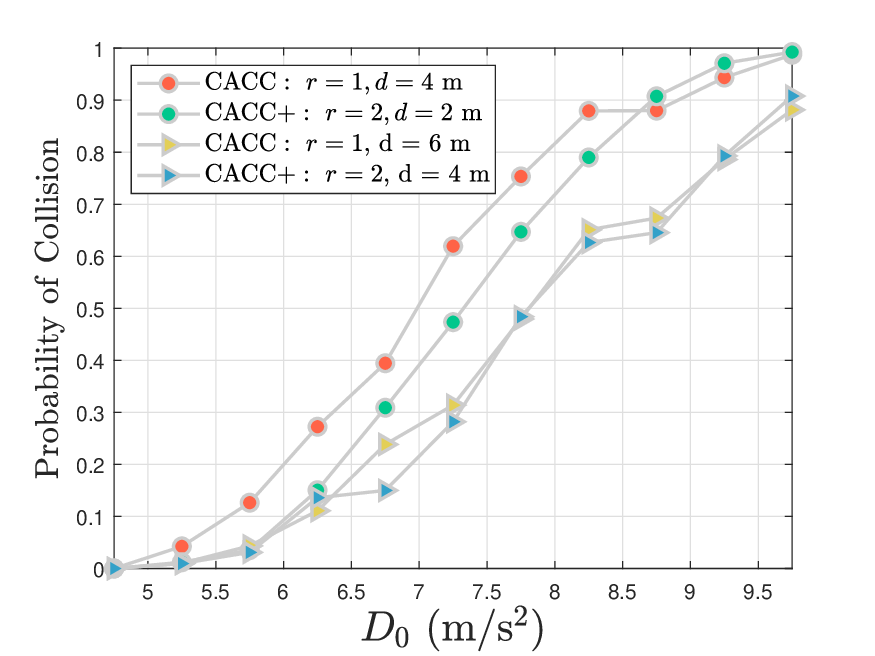}}
\caption{The benefit of CACC+ in terms of probability of collision.
\label{fig:P_benefit}}
\end{figure}

\begin{figure}[!htb]
\centering{\includegraphics[scale=0.55]{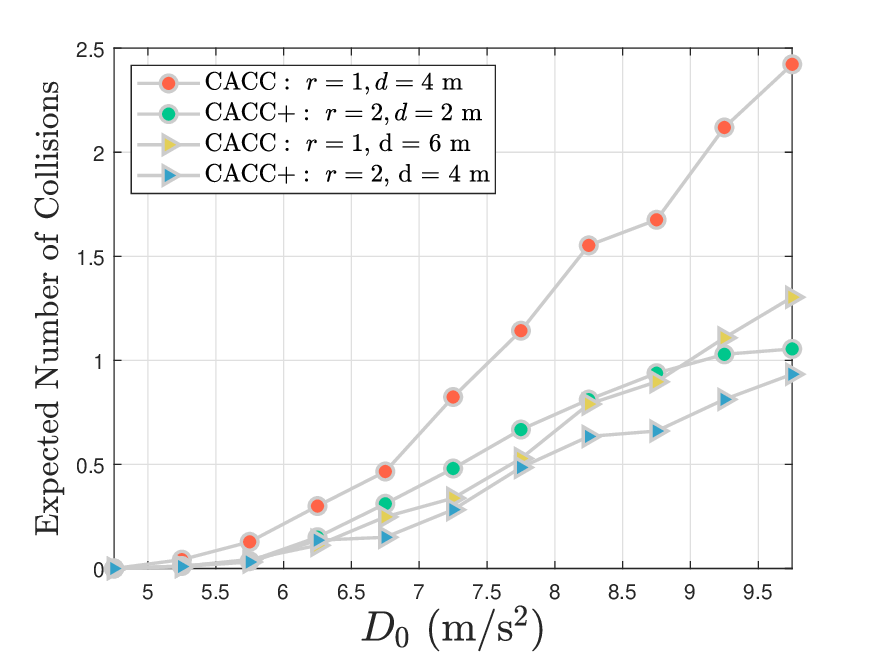}}
\caption{The benefit of CACC+ in terms of the expected number of collisions.
\label{fig:N_benefit}}
\end{figure}

While the safety metrics can be improved by increasing the standstill spacing, however, with the increase of the standstill spacing, the vehicle throughput and connectivity reduce. To evaluate the effectiveness of CACC+ in improving the safety metrics, the comparison of the probability of collision between the CACC case with $d = 4$ m and $d = 6$ m and the CACC+ case ($r = 2$) with $d = 2$ m and $d = 4$ m is provided in Fig.~\ref{fig:P_benefit}, and the comparison of the expected number of collisions between the CACC case with $d = 4$ m and $d = 6$ m and the CACC+ case ($r = 2$) with $d = 2$ m and $d = 4$ m is provided in Fig.~\ref{fig:N_benefit}. It can be seen that CACC+ can achieve comparable or better performance with a smaller standstill spacing when compared to CACC. 

\vspace*{-0.1in}
\section{CONCLUSION} \label{section:conclusion}
In this paper, we have provided a new approach to assess safety benefits of cooperative adaptive cruise control systems with different information topologies in connected and autonomous vehicles. We have compared CACC and CACC+ under emergency braking scenarios with safety metrics given by the probability of collision, the expected number of collisions, and the relative velocity at impact. By varying the standstill spacing, we found larger standstill spacing can render better safety performance, and using CACC+ with a smaller standstill spacing can maintain a comparable safety performance than CACC. Topics of future work include safety implications when there are packet losses in communicated information, communication delay, noise, effect of control gains and time headway. 

\vspace*{-0.1in}
\bibliographystyle{IEEEtran}

\bibliography{ref}

\end{document}